# Spatiotemporal helicon wavepackets

PIERRE BEJOT[1,*] AND BERTRAND KIBLER[1]

[1]*Laboratoire Interdisciplinaire Carnot de Bourgogne, UMR 6303 CNRS / Université Bourgogne Franche-Comté, 21078 Dijon, France*
*\*pierre.bejot@u-bourgogne.fr*

**Abstract:** Propagation-invariant or non-diffracting optical beams have received considerable attention during the last two decades. However, the pulsed nature of light waves and the structured property of optical media like waveguides are often overlooked. We here present a four-dimensional spatiotemporal approach that extends and unifies both concepts of conical waves and helicon beams, mainly studied in bulk media. By taking advantage of tight correlations between the spatial modes, the topological charges, and the frequencies embedded in an optical field, we reveal propagation-invariant (dispersion- and diffraction-free) space-time wavepackets carrying orbital angular momentum (OAM) that evolve on spiraling trajectories in both time and space in bulk media or multimode fibers. Besides their intrinsic linear nature, we show that such wave structures can spontaneously emerge when a rather intense ultrashort pulse propagates nonlinearly in OAM modes. With emerging technologies of pulse/beam shaping, multimode fibers and modal multiplexing, our proposed scheme to create OAM-carrying helicon wavepackets could find a plethora of applications.

## 1. Introduction

Light structuring has found far-reaching applications covering numerous fields such as fundamental physics [1], telecommunications [2-3], particle manipulation [4-5], quantum information technology [6-7], laser micromachining [8-9], and high-resolution imaging [10-13]. Propagation-invariant or non-diffracting beams belong to the important class of structured optical fields that has received a great deal of attention [14-15]. Propagation-invariant beams are particular solutions of Maxwell's equations that have the particularity to sustain their shape all along the propagation. The simplest well-known (theoretical) diffraction-free beams are monochromatic plane waves which are solutions of Maxwell's equation in free space in Cartesian coordinates. Apart this trivial solution, monochromatic invariant beams have been widely studied in free space, both theoretically and experimentally since several decades. For instance, one can cite Bessel [14], Mathieu [16], Weber [17], and self-accelerating Airy beams [18], which are solutions of free space Maxwell equations in cylindrical, elliptic, or parabolic accelerating coordinates.

Orbital angular momentum (OAM)-carrying beams (i.e., vortex beams) are also solutions of Maxwell's equations and have drawn increasing interest in recent years [19-20]. They are characterized by a phase singularity at the center, a helical wavefront, and their so-called topological charge, an integer number counting how many wavefront rotations take place over one propagation wavelength. Even if it is not a general law for such beams, some of them are also diffraction-free. This is in particular the case for higher-order Bessel beams. This is important to note that, even if OAM beams own helical wavefronts and that the direction of their Poynting vectors rotates around the propagation axis, their intensity patterns do not. Indeed, their donut-like intensity pattern being cylindrically symmetric, they cannot exhibit any rotation. However, particular superposition of diffraction-free Bessel beams that carry OAM can provide rotating complex beams with spiraling (or screw-shaped) trajectories of their intensity distribution during propagation like helicon beams [21-23]. More complex forms of OAM-carrying beams have been also investigated numerically such as rotating-revolving Laguerre-Gaussian beams using multiple optical-frequency-comb lines [24]. More generally, by carefully controlling the interference of structured light that caries OAM, twisting and accelerating light over both the radial and the angular directions can be demonstrated [25-26], as well as possibly providing an optical pulling force through tractor beams [27-28].

On the other hand, propagation-invariant *pulsed* beams (without carrying OAM) have also been intensively studied. Such solutions are not only diffraction-free as their monochromatic counterpart but are also insensitive to dispersion, i.e., they do not spread in time as they propagate in dispersive bulk media. Brittingham's focus-wave modes [29], Mackinnon's wave packet [30], Bessel-pulse X-pulses [31], and space-time light sheets [32] belong to this particular class of propagation-invariant wavepackets. The non-dispersive and non-diffractive nature of such fully localized (in space and time) waves is due to their intrinsic coupled spatiotemporal properties by which the combined effects of spatial diffraction and temporal dispersion cancel each other. Moreover, it has been shown that X-waves can also spontaneously emerge when intense ultrashort pulses nonlinearly propagate in dispersive bulk media [33].

As far as structured media (i.e., waveguides) are concerned, the monochromatic diffraction-free beams are known since a long time and are called optical modes of the waveguide. Recently, a generalization of propagation-invariant conical waves in dispersive waveguides, called discretized conical waves, has been derived [34]. As their bulk counterpart, such wavepackets do not spread in space and time. Moreover, it has also been shown that such waves are naturally generated when a rather intense ultrashort pulse having a cylindrical symmetry around the propagation axis nonlinearly propagates in the waveguide. The intrinsic difference between discretized and bulk conical waves comes from the discrete number of modes that can propagate in a waveguide, so that their spectrum is also discrete.

When one now considers all the concepts above listed, namely the propagation-invariance, the pulsed nature, the OAM features of light waves and the structured property of optical media like waveguides, it appears essential to establish a global framework able to account for their simultaneous combination in both bulk media and waveguides for future developments and applications of light structuring. To this end, in this article, we present a novel class of invariant spatiotemporal wavepackets guided in multimode optical fibers, namely OAM-carrying helicon wavepackets. Note that our generalized approach is also valid for bulk media, since the latter can be considered as a waveguide of infinite dimension. The main feature of such dispersion- and diffraction-free electric fields is that they rotate in both time and space as they propagate, without any space-time distortion, at a chosen group-velocity in the fiber core. Such an amazing property is a consequence of tight correlations between the spatial modes, the topological charges, and the frequencies embedded in the optical field. Beyond the possible linear shaping of such structured waves, we also reveal their spontaneous emergence triggered by the nonlinear propagation of a rather intense ultrashort pulse in fiber OAM modes. In particular, we will show that such wavepackets, spiraling in both time and space, are linked to the formation of helical optical shocks.

The article is organized as follows. We first recall the theoretical description of OAM modes and then provide the construction of linear spatiotemporal helicon wavepackets based on a strict multidimensional phase-matching condition. A simple analytic example is then presented and discussed in the case of a fused silica rod (i.e., at the frontier between bulk media and fibers). Next we numerically investigate the nonlinear generation of such wave structures through the use of a multimode fiber and its OAM modes pumped by an intense ultrashort pulse. In particular, we confirm that the formalism of helicon wavepackets introduced in the first section allows us to predict accurately the mode-resolved output spectrum of the laser pulse at the output of the multimode fiber.

## 2. Theoretical description: spatiotemporal helicon wavepackets

*3.1 Basic considerations*

This section is devoted to the theoretical description of propagation-invariant (dispersion- and diffraction-free) space-time wavepackets that evolve on spiraling trajectories around the propagation axis in structured/finite media. For simplicity, we will restrict our study to a scalar approach by considering the weak guidance approximation. Moreover, we will study only waveguides whose refractive index presents a cylindrical symmetry around the propagation axis $z$ (i.e., standard optical fibers). Using the above restriction, the partial differential equation driving the linear propagation of an electric field $E$ is given in cylindrical coordinates by:

$$\left[\partial_z^2 + \partial_r^2 + \frac{1}{r}\partial_r + \frac{1}{r^2}\partial_\theta^2 + \frac{n^2(r,\omega)\omega^2}{c^2}\right]\tilde{E}(r,\theta,\omega) = \Box\tilde{E}(r,\theta,\omega) = 0, \quad (1)$$

where $c$ is the light velocity in vacuum, $n(r,\omega)$ is the radial-dependent refractive index at the angular frequency $\omega$, and $\tilde{E}$ is the Fourier transform of $E$ with respect to the time coordinate:

$$\tilde{E}(\omega) = \int E(t)e^{i\omega t}dt. \quad (2)$$

Before going ahead, let us find solutions of Eq. 1 that write

$$\tilde{\varepsilon}(r,\theta,z,\omega) = \tilde{A}(r,\theta,\omega)e^{iK_z z}. \quad (3)$$

Such solutions are called optical modes of the fiber and $K_z$ is called the propagation constant of the mode. Note that modes are not necessarily propagating but can be exponentially decaying along $z$ provided that $K_z$ is complex. Moreover, note that, in the context of optical fibers, a mode can propagate in the fiber cladding, even if it is of common use to call "modes" only the ones confined within the core. Since the electric field is intrinsically $2\pi$-periodic (and assumed piecewise-continuous) with respect to $\theta$, its angular dependence can be developed as a Fourier-series:

$$\tilde{A}(r,\theta,\omega) = \sum_l \widetilde{A_l}(r,\omega)e^{il\theta}. \quad (4)$$

By injecting Eqs. 3 and 4 in Eq. 1, one obtains:

$$\sum_l \left[\partial_r^2 + \frac{1}{r}\partial_r - \frac{l^2}{r^2} + \frac{n^2(r,\omega)\omega^2}{c^2} - K_z^2\right]\widetilde{A_l}(r,\omega)e^{i(l\theta+K_z z)} = \sum_l \Box_l \widetilde{A_l}(r,\omega)e^{i(l\theta+K_z z)} = 0. \quad (5)$$

Since the refractive index is assumed here to be angularly invariant, in bra-ket notation, one has:

$$\langle e^{il'\theta}|\Box_l\widetilde{A_l}(r,\omega)e^{i(l\theta+K_z z)}\rangle = \Box_l\widetilde{A_l}(r,\omega)e^{iK_z z}\delta_{l,l'}, \quad (6)$$

where $\delta_{l,l'}$ is the Kronecker delta. It then indicates that optical modes of such fibers are necessarily of the form:

$$\tilde{\varepsilon}(r,\theta,z,\omega) = \tilde{A}(r,\omega)e^{i(l\theta+K_z z)}. \quad (7)$$

In other words, functions $|e^{il\theta}\rangle$ describe the angular part of the optical modes, and this means that OAM modes form a basis set to represent spatial modes. This is not true anymore for a fiber whose refractive index depends on the angular coordinate $\theta$, since, in this case, the operator $\Box_l$ will couple the different functions $|e^{il\theta}\rangle$. Equation 7 gives the functional form of the optical modes in an angularly-invariant waveguide. Moreover, one will impose that the electric field vanishes at a given radius $R$, which is located sufficiently far from the waveguide:

$$\tilde{A}(r = R, \omega) = 0. \quad (8)$$

Imposing such a Dirichlet boundary condition then discretizes the number of eligible solutions. Consequently, for a given couple $(l, \omega)$, the solutions (i.e., modes) can be indexed and sorted by an integer number $p = 1 \cdots \infty$, the radial order related to the number of concentric intensity rings in the intensity profile. As a consequence, a fiber mode can be represented by a triplet $(l, p, \omega)$. Moreover, after an appropriate normalization, the modes form an orthonormal basis so that any electric field $E$ can be expressed as

$$E(r, \theta, t) = \int \sum_{l,p} \bar{E}(l, p, \omega) A_{l,p}(r, \omega) e^{i(l\theta - \phantom{x})} d\omega, \qquad (9)$$

where $A_{l,p}(r, \omega)$ is the transverse envelope of the mode $(l, p, \omega)$ and $\bar{E}(l, p, \omega)$ are the electric field coordinates in the modal basis.

Using the unidirectional pulse propagation equation (UPPE) [35], the linear evolution of an electric field $E(r, \theta, z, \omega)$ along the propagation axis $z$ is given in the modal basis by

$$\partial_z \bar{E} = i K_z(l, p, \omega) \bar{E}, \qquad (10)$$

where $K_z(l, p, \omega)$ is the propagation constant of the OAM mode $(l, p)$ at the frequency $\omega$. Recall that $l$ and $p$ refer here to azimuthal and radial indices, where $l(0, \pm 1, \pm 2, \pm 3, \ldots)$ is the topological charge, related to the phase front of the OAM mode. OAM modes are higher-order modes defined on a different basis as compared to more conventional modes in fiber, such as linearly polarized (LP) modes and vector modes, so that they can be also regarded as the linear combination of the latter [36-37]. For example, we can here write $A_{l=\pm 1, p=1} = (\mathrm{LP}_{11}^{even} \pm i \mathrm{LP}_{11}^{odd})/\sqrt{2}$, as well as $A_{0,1}$ corresponds to the fundamental mode $\mathrm{LP}_{0,1}$ of the fiber. The $+i$ or $-i$ terms represent a $\pi/2$ or $-\pi/2$ phase shift in the linear combinations.

*3.2 Definition of helicon wavepackets*

One can now define families of modes $(l, p, \omega_{lp})$ in such a way that their propagation constants all verify the same relation:

$$K_z(l, p, \omega_{lp}) = K_0 + K_1 \omega_{lp} + K_l l, \qquad (11)$$

where $K_0$, $K_1$, and $K_l$ are arbitrarily chosen constants. Note that, depending on the chosen constants defining a family, the latter can contain *a priori* several frequencies in the same spatial mode and a given frequency can satisfy relation (10) in different spatial modes. A corresponding electric field can be built from the linear superposition of the modes belonging to the same family:

$$E(r, \theta, t) = \sum_{l,p} \bar{E}(l, p, \omega_{lp}) A_{l,p}(r, \omega_{lp}) e^{i(l\theta - \omega_{lp} t)}. \qquad (12)$$

Using Eqs. 10-11, one then obtains the evolution of the above electric field:

$$E(r, \theta, z, t) = E(r, \phi, z = 0, \tau) e^{i K_0 z}, \qquad (13)$$

where $\phi = \theta + K_l z$ and $\tau = t - K_1 z$. Accordingly, any electric field built from a given family is a diffraction- and dispersion-free space-time wavepacket propagating at the group velocity $1/K_1$ whose intensity continuously rotates around the propagation axis with a spatial period $2\pi/K_l$. In other words, the built families are perfectly invariant fields in a frame propagating at the velocity $1/K_1$ and rotating around the $z$ axis with a period $2\pi/K_l$. Wavepackets with spiraling trajectories then result from the linear superposition of waves oscillating at different frequencies and carrying different topological charges.

Several remarks have to be made concerning such wave structures. First, the fact that the spatiotemporal intensity rotates around the propagation axis does not necessarily imply that the fluence (i.e., the energy per surface unit) does. In fact, using the Parseval theorem, one can show that the fluence exhibits a rotation around the propagation axis only if the defined family embeds at least two modes $(l_1, p_1, \omega_{l_1,p_1})$ and $(l_2, p_2, \omega_{l_2,p_2})$ such that $\omega_{l_1,p_1} = \omega_{l_2,p_2}$ with $l_1 \neq l_2$. Otherwise, the fluence of the electric field presents a cylindrical symmetry around the propagation axis so that no rotation of the fluence is observed. Reciprocally, the instantaneous power of the wavepacket (i.e., the total energy per time unit) is constant except if the electric field embeds at least two distinct frequencies in the same spatial mode.

The above construction of wavepackets, composed of discrete frequencies, implies an infinite energy, which cannot correspond to a realistic physical situation, except if one uses a set of continuous wave lasers (or multiple frequency-comb lines) combined with spatial shaping and multiplexing. On the other hand, the carrier frequencies $\omega_{lp}$ involved could have a finite narrow bandwidth (i.e., associated with a temporal envelope $a_{l,p}$). As a consequence, the electric field constructed from a given family would write in good approximation as

$$E(r, \theta, t) = \sum_{l,p} \bar{E}_{l,p,\omega_{lp}} a_{l,p}(t) A_{l,p}(r, \omega_{lp}) e^{i(l\theta - \omega_{lp} t)}, \tag{14}$$

so that its evolution along the propagation axis is

$$E(r, \phi, z, \tau) = \sum_{l,p} \bar{E}_{l,p,\omega_{lp}} a_{l,p}\left(\tau - \Delta K_{z,lp}^{(1)} z\right) A_{l,p}(r, \omega_{lp}) e^{i(l\phi - \omega_{lp} \tau)}, \tag{15}$$

where $\Delta K_{z,lp}^{(1)} = K_{z,lp}^{(1)} - K_1$ with $K_{z,lp}^{(1)} = \{\partial_\omega K_z(l, p, \omega)\}_{\omega = \omega_{lp}}$ the inverse of the group velocity of the mode $(l, p)$ taken at the frequency $\omega_{lp}$. Accordingly, a wavepacket embedding a finite amount of energy is necessarily dispersive, i.e., not strictly speaking dispersion- and diffraction-free. Note however, that the effect of dispersion, i.e., the length from which the dispersion of the group velocities in the different modes will impact the propagation, will strongly depend on the envelope duration. Then, the wavepacket can be considered as invariant but over a finite distance.

Last but not least, if now, one comes back to the essential issue of establishing a general framework able to link the concepts of propagation-invariance, pulsed nature, and OAM, it is important to notice that modes of a given family lye on a surface *S* in the three-dimensional (*l,p,ω*) modal space. This surface embeds both helicon (monochromatic) beams and discretized conical wavepackets [34], which are located at the intersection between *S* and planes of constant *ω* and constant topological charge *l*, respectively. Our theory then naturally encompasses both concepts which now appear as special cases of spatiotemporal helicon wavepackets.

### 3.3 Simple example of helicon wavepackets

Before studying how the above wavepackets can emerge from nonlinear propagation in a standard multimode fiber, we first construct such space-time wave structures in a simple case at the frontier between bulk media and fibers. In particular, we consider a dispersive medium of finite transversal dimension (radius *R*), often used as a toy model for calculating the guided modes of hollow core capillaries. Its modes have the advantage to be perfectly analytic as follows:

$$\tilde{\varepsilon}_{l,p}(r, \theta, z, \omega) = J_l\left(\frac{\alpha_{lp} r}{R}\right) e^{i(K_z(\omega) z + l\theta)}, \tag{16}$$

where $\alpha_{lp}$ is the $p^{th}$ root of $l^{th}$ Bessel function of first kind $J_l$, $K_z(\omega) = \sqrt{k^2(\omega) - \alpha_{lp}^2/R^2}$, $k(\omega) = n(\omega)\omega/c$ and $n(\omega)$ is the frequency-dependent refractive index of the medium. In the paraxial approximation, the propagation constants can be approximated by

$$K_z(l,p,\omega) \simeq k(\omega) - \frac{\alpha_{lp}^2}{2k(\omega)R^2}. \tag{17}$$

Note that such an approximation does not change the underlying physics but it allows to deal with more tractable analytical formula. Moreover, we assume that the wavevector $k(\omega)$ can be developed in frequency as a Taylor series around a central carrier frequency $\omega_0$. Accordingly, one has

$$K_z(l,p,\Omega) \simeq \sum_n \frac{k_n}{n!}\Omega^n - \frac{\alpha_{lp}^2}{2k_0 R^2}, \tag{18}$$

where $\Omega = \omega - \omega_0$ and $k_n = \{\partial_\omega^n k\}_{\omega=\omega_0}$. We can then define the family of modes that satisfies:

$$K_z(l,p,\Omega_{lp}) \simeq \sum_n \frac{k_n}{n!}\Omega_{lp}^n - \frac{\alpha_{lp}^2}{2k_0 R^2} = K_0 + K_1\Omega_{lp} + K_l l. \tag{19}$$

As a result, for each couple $(l,p)$, one has to find the roots of a $n^{th}$ order polynomial in $\Omega_{lp}$. The purely real roots then correspond to frequencies that satisfy Eq. 19. As an example, Figure 1(a) depicts the family of modes given by the above condition with $K_l = K_z(1,1,0) - K_z(0,1,0)$, when considering a fused silica rod ($R = 100$ μm) and a central wavelength $\lambda_0 = 2\pi c/\omega_0 = 800$ nm. Note that we added an arbitrary group delay of $\delta k_1 = 3$ ps m$^{-1}$, i.e., the group velocity of the helicon wavepacket is $1/K_1 = 1/(k_1 + \delta k_1)$. Then, we observe an evident combination of X-shaped patterns whose individual element is similar to the well-known X-shaped conical wave studied in previous works [33-34].

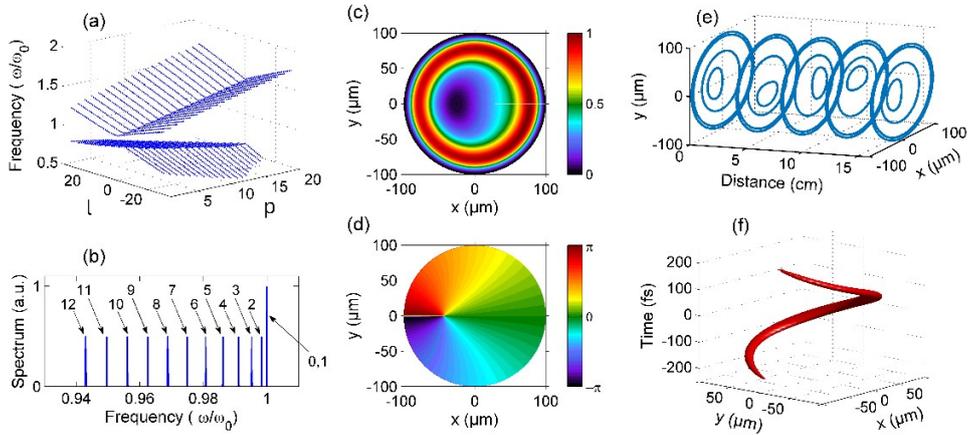

**Fig. 1.** Linear construction of a spatiotemporal helicon wavepacket in a fused silica rod. (a) Family of modes satisfying the phase-matching condition (19) for the following parameters: $\lambda_0 = 800$ nm, and $\delta k_1 = 3$ ps m$^{-1}$. (b) Power spectrum of the few selected modes involved in our linear superposition. (c) Corresponding initial fluence distribution in space. (d) Corresponding phase distribution at $\omega_0$. (e) Evolution of iso-contour of the fluence with propagation distance (at $1/e^2$ and half-maximum) showing the counterclockwise rotation around the propagation axis. (f) Corresponding iso-surface of the spatiotemporal intensity profile at half-maximum that features the localized helicon wavepacket.

Indeed, the overall 3D-pattern of phase-matching results from the combination of all phase-matched frequencies from modes with various radial and angular indices, which then creates discretized conical waves over $p$- or $l$- dimension according to the dispersion regime. If one considers a few modes characterized by low radial and angular indices to construct an example of helicon wavepackets, a simple solution can be written by considering the second-order dispersion only:

$$\Omega_{lp,\pm} = (K_1 - k_1 \pm \sqrt{\Delta})/k_2 \tag{20}$$

with $\Delta = (K_1 - k_1)^2 - 2k_2 \left[ k_0 - K_0 - \frac{\alpha_{lp}^2}{2k_0 R^2} - K_l l \right]$. Figure 1(b) depicts a set of 12 modes in addition to the fundamental fiber mode given by the relation:

$$K_z(l, p, \Omega_{lp}) = K_z(0,1,0) + (k_1 + \delta k_1)\Omega_{lp} + K_l l \tag{21}$$

The linear construction of the helicon wavepacket then results from the superposition of frequencies $\Omega_{l=0\ldots12, p=1,-}$ with equal spectral amplitudes and phases (see Fig. 1(b-d)). Since the present example corresponds to the particular case, where two distinct angular modes oscillates at the same frequency (here, $\Omega_{01,-} = \Omega_{11,-} = 0$), the fluence distribution is angularly asymmetric with zero-value shifted from the origin (i.e., distinct from a perfect donut-like pattern), as shown in Fig. 1(c). The spatial phase distribution at $\omega_0$ is shown in Fig. 1(d), the complex pattern results from the superposition of two OAM fiber modes and is characterized by a singularity shifted from the origin. The spatial phase distribution for other frequencies that contain a single mode is not shown here since featured by typical signatures related to the topological charge (i.e., a phase singularity at the center and $l$ times $2\pi$-azimuthal phase changes).

The full electric field at a given propagation distance $z$ is given by Eqs. 12-13. Figure 1(e) shows the evolution of the fluence along propagation which implies the spiraling trajectories of both fluence and phase profiles while maintaining their initial distribution. The rotation period is directly given by $2\pi/|K_l| \simeq 16$ cm, and the direction of rotation is counter clockwise as driven by the negative value of $K_l$. As shown in Fig. 1(f) that displays the iso-surface of the spatiotemporal intensity at half maximum, the wavepacket looks like a corkscrew in space-time coordinates, the points of high intensity being extremely localized in both space and time since all frequencies are in phase. Nevertheless, all modes used for the construction contain a single frequency so that the instantaneous power (integrated over spatial coordinates) is constant. Note that despite the time-localization of our helicon wavepacket, this ultrashort structure does not disperse and simply rotates during its propagation around the $z$- and $t$-axis because of the inherent invariant nature of the constructed wavepacket (see supplementary movie).

Note that more complex fluence and phase profiles could be easily generated by choosing a superposition of multiple higher-order modes at the phase-matched discrete frequencies. This would provide simultaneous revolving and rotating of such profiles in the x–y plane during propagation [24].

## 3. Numerical simulations: nonlinear generation in multimode fibers

Multimode fibers are known as versatile platforms, in particular, for sculpting supercontinuum as a function of the initial pump spatial profile [38-40]. In the following, we show how spatiotemporal helicon wavepackets can be nonlinearly generated by appropriate and realistic initial pumping conditions. Our numerical approach of nonlinear pulse propagation is based on the multimode unidirectional pulse propagation equation recently

derived in [41], which describes the evolution of the complex electric field in the scalar approximation:

$$\partial_z \bar{E} = i\left[K_z - \frac{\omega}{v_{g_0}}\right]\bar{E} + \frac{i\omega^2}{2\epsilon_0 c^2 K_z}\bar{P}_{NL}, \quad (22)$$

$\epsilon_0$ is the vacuum permittivity, and $\bar{P}_{NL}$ is the nonlinear polarization expressed in the modal basis. Moreover, for convenience, the equation is written in a local frame propagating at an arbitrarily chosen velocity $v_{g_0}$. The latter is often chosen as the group velocity of the fundamental mode ($l = 0, p = 1$) calculated at $\omega_0$. In the context of fiber propagation, using a complex representation of the electric field $\xi$ (expressed now so that $|\xi|^2 = I(r, \theta, t)$, $I$ being the pulse intensity), UPPE rewrites:

$$\partial_z \bar{\xi} = i\left[K_z - \frac{\omega}{v_{g_0}}\right]\bar{\xi} + \frac{in_{\text{eff}_0} n_2 \omega^2}{c^2 K_z}\left\{(1-f_R)\overline{|\xi|^2 \xi} + f_R \overline{[\int h_R(\tau)|\xi(t-\tau)|^2 d\tau]\xi}\right\}, \quad (23)$$

where $n_{\text{eff}_0}$ is the effective refractive index of the fundamental mode at $\omega_0$, $n_2$ is the nonlinear refractive index of the medium (here for silica glass, we used $n_2 = 3.2\ 10^{-20}$ m$^2$/W). The function $h_R$ is the Raman response with fraction $f_R = 0.18$ for fused silica glass. For simplicity, we have neglected here the nonlinear term responsible for third harmonic generation. We recall that the present simulations are full 3D+1 simulations. Note that the effect of usual fiber losses (less than 10 dB/km) is found here to be negligible over the considered propagation distance. We solve the propagation by a split-step algorithm as described in Ref. [41]. Counting both radial and angular modes, we considered more than 1500 modes for our calculations. Using such a number of modes would be completely intractable if an algorithm based on multimode generalized nonlinear Schrodinger equation (MM-GNLSE) were employed [42].

As an example (see Fig. 2), we investigate the nonlinear propagation of 100-fs Gaussian pulse at $\lambda_0 = 1300$ nm, with 450-nJ energy, close to the zero dispersion of a standard step-index multimode fiber (core radius $R = 52.5$ μm and numerical aperture NA = 0.22). The associated peak power is around 0.8 times the critical power of silica glass. The energy is initially equally-distributed on the two first OAM modes of the fiber ($p = 0, l = 0$ and $+1$). Note that such an initial condition could be experimentally realized using a spatial beam shaper. The propagation constant difference between the OAM modes at $\omega_0$ is $K_z(1,1,0) - K_z(0,1,0) \simeq$ -220 rad m$^{-1}$. Accordingly, the laser pulse initially rotates around the propagation axis with a period $\sim 2.8$ cm, as shown in Fig. 2(c). We observe about 3 rotations of the guided beam in the fiber core (denoted by the contour-plot of fluence at half-maximum) over the 8.5-cm-long propagation distance under study. The detailed nonlinear propagation for temporal power and power spectrum is shown in Fig. 2(a-b), respectively. We observe that the spectral dynamics reaches a stationary state after 7 cm (see Fig. 2(b)), just past the pulse splitting dynamics observed in the time domain. The initial self-focusing dynamics arrested by higher order dispersions and self-steepening leads to the spontaneous formation of localized sub-pulses (at the trailing edge) in the time domain and a strong spectral broadening. Sub-pulses are moving in the same direction in the retarded time frame with nearly the same group velocity (different from $v_{g_0}$) as indicated by the white dashed arrow in Fig. 2(a), and characterized by a relative group delay of 800 fs m$^{-1}$. The simultaneous strong broadening associated to the most intense sub-pulse seeds linear waves, which are resonantly amplified in higher order modes. A clear structure emerges in the mode-resolved spectrum at the distance of occurrence of this sub-pulse, in the form of discretized conical fish-wave (also known as a combination of both X-wave and O-wave) over angular indices $l$ for low radial indices $p$, as depicted in Fig. 2(d). This particular shape corresponds to the formation of a spatiotemporal helicon wavepacket belonging to a particular family of modes as defined in the previous section by Eqs. 11 and 12.

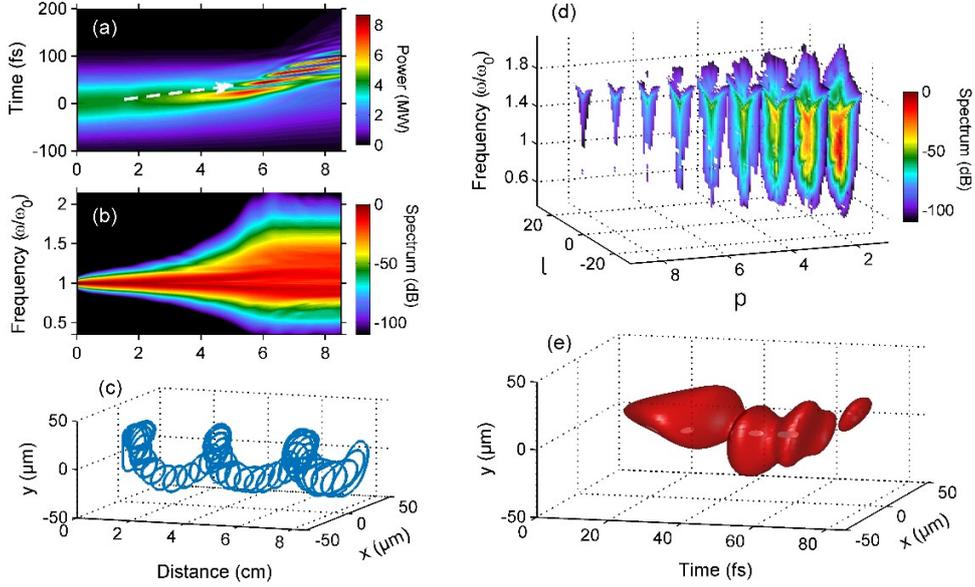

**Fig. 2.** Nonlinear propagation of 100-fs pulses (450-nJ energy at 1300 nm) in a 8.5-cm-long segment of step-index multimode fiber. (a)-(b) Evolution of the instantaneous power and normalized full power spectrum with propagation distance, respectively. The white dashed arrow in panel (a) indicates the group-velocity of the most intense sub-pulse. (c) Evolution of iso-contour at half-maximum of the fluence with propagation distance. (d) Output mode-resolved power spectrum as a function of $p$ and $l$ indices observed after 6.1 cm of propagation. (e) Corresponding iso-surface at $1/e^2$ of the 3D-intensity profile.

In other words, this simply corresponds to the linear superposition of phase-matched resonant radiations over the fiber OAM modes. More particularly, the emerging helicon wavepacket propagates and rotates according to the velocity features of the most intense sub-pulse.

Based on the three-wave mixing (TWM) picture [43], previously developed for describing the generation of conical waves in bulk media, the underlying basic idea is here to consider that the OAM-carrying supercontinuum generated during the nonlinear propagation scatters on a "material wave" which represents the nonlinearly-induced change of the refractive index of the medium. While previous conical waves in bulk media and multimode fibers are solely characterized by their propagation velocity (see Refs. [34,43]), it is here defined by two parameters, namely its propagation velocity and rotation period due to their spiraling trajectory based on OAM fiber modes. Accordingly, the energy of the scattered waves will be concentrated in families of modes $(l, p, \omega_{lp})$ propagating and rotating at the same rates than the material wave (MW):

$$\left| K_z(l, p, \omega_{lp}) - K_{0,\text{MW}} - K_{1,\text{MW}} \omega_{lp} - K_{l,\text{MW}} l \right| \leq \frac{2\pi}{d_r}, \qquad (24)$$

where $K_{0,\text{MW}}$, $K_{1,\text{MW}}$, and $K_{l,\text{MW}}$ are the propagation constant, the inverse of group velocity, and the rotation rate of the material wave, respectively. Here, the material wave corresponds to the most intense sub-pulse that is triggered and that swirls around the propagation axis. Since the nonlinear process takes place over a finite length $d_r$, Eq. 24 is not a strict phase-matching condition, it can be satisfied over a certain tolerance related to $d_r$ [34,43]. Equation

24, together with the precise knowledge of the frequency-dependent propagation constants, then allows to estimate the mode-resolved output spectrum shape of the pulse impacted by the conical wave emission. $K_{0,MW}$ is equal to the propagation constant of the input pump pulse $K_0(l = 0, p = 1, \omega_0)$. From Fig. 2(a) and the white dashed arrow, the inverse of group velocity is found to be $K_{1,MW} = 1/v_{g_0} + \delta k_1$, with $\delta k_1 = 800$ fs m$^{-1}$.

Concerning the rotation rate $K_{l,MW}$, one can expect that it is also impacted by the nonlinear propagation in a similar way to the group velocity. This is corroborated by the 3D-intensity profile spiraling in time and obtained at the distance of conical wave emission, and depicted in Fig. 2(e). The iso-surface of the intensity clearly exhibits distinct sub-pulses spiraling, characterized by different positions and shapes in space, thus suggesting a variation of rotation feature along the pulse splitting. To better unveil this effect, we show the corresponding cross-sections of spatial intensity profiles at distinct times in Fig. 3(a-c), namely the center of our retarded time frame, and the temporal locations of the maximal intensity of the two first sub-pulses, respectively. As above suggested, we clearly observe that the intensity peaks do not rotate at the same velocity as the initial propagation constant difference between the OAM modes at $\omega_0$. The angular delay between the most intense sub-pulse (Fig. 2(d)) and the linear rotation at $\omega_0$ is found to be ~1.46 rad after 6.1 cm. As a consequence, the rotation rate of the material wave can be determined as $K_{l,MW} = K_z(1,1,0) - K_z(0,1,0) + \delta k_l$, with $\delta k_l \simeq 24$ rad m$^{-1}$. From the above determination of $K_{1,MW}$ and $K_{l,MW}$, we can now provide a detailed analysis of the 3D-phase-matching pattern depicted in Fig. 2(e) by using our theoretical approach, in order to confirm the emission of the helicon wavepacket.

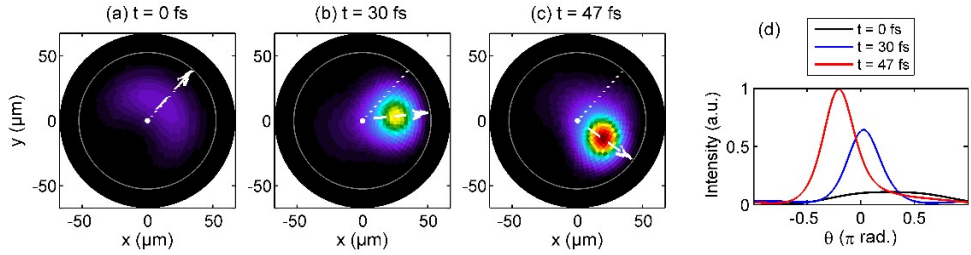

**Fig. 3.** Cross-sections of 2D-spatial intensity profiles studied in Fig. 2(e) at distinct times of our retarded frame, namely (a) $t = 0$ fs, (b) $t = 30$ fs, and (c) $t = 47$ fs. The gray circle corresponds to the core/cladding interface. The white dotted line is the angular position from $\omega_0$ (~13.4 rad from the input, i.e. 2.13 rotations) whereas the white dashed arrow indicates the angular position of the maximal intensity. (d) Intensity profiles along the angular position for the different times under study (note that the radial positions of maximal intensity were chosen).

Figure 4 shows the angularly-resolved spectra for the five first radial and angular indices obtained from numerical simulation at the output of the fiber (i.e., after 8.5 cm of propagation). The white iso-contour superimposed to the mode-resolved output spectrum delimits the inner angular-spectral region that satisfies Eq. 24 with parameters fixed above when $d_r = 5$ mm (roughly the distance corresponding to the lifetime of the most intense sub-pulse). All the mode-resolved output spectral shapes are well fitted by this contour, which confirms that a helicon wavepacket rotating at $K_{l,MW}$ and propagating at group velocity $1/K_{1,MW}$ is generated during the nonlinear propagation. A clear conical wave of fish-wave type is revealed over the lowest orders of OAM fiber modes, a typical signature of pumping close to the zero dispersion of the medium or waveguide [34,43]. We also notice that the bandwidth of phase-matching condition strongly reduces for increasing values of radial or angular indices (i.e., higher order modes). Note that, the tolerance of phase-matching condition in Eq. 24 (as the wavepacket is generated over a finite distance) makes that spectral components forming the wavepacket are not delta functions distributed over a given family of modes, in contrast with purely non-spreading wavepackets. Each mode composing the helicon wavepacket here supports a certain bandwidth, thus the latter will be dispersive, as described by Eq. 15.

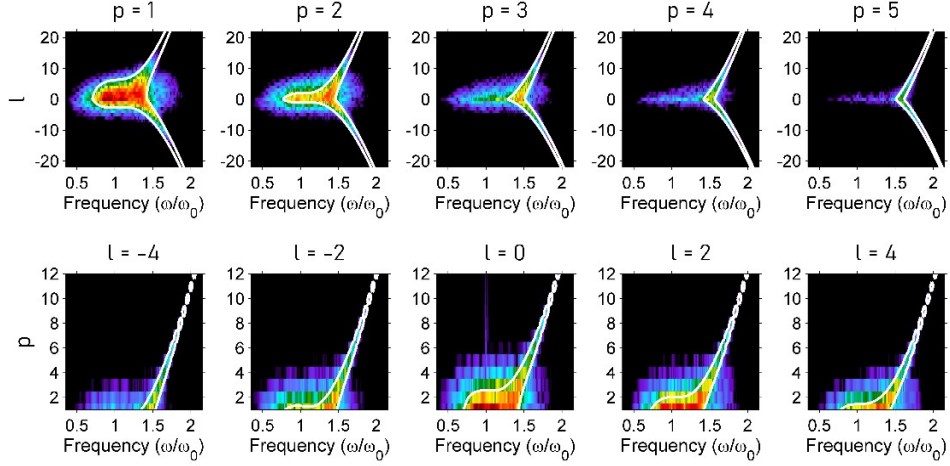

**Fig. 4.** Angular-mode-resolved output spectra for the first radial (top panels) and angular (bottom panels) indices. The white contour corresponds to the phase-matching of theoretical radiations forming the spatiotemporal helicon wavepacket propagating with a group delay $\delta k_1$ and rotating around the propagation axis with angular delay $\delta k_l$ compared to the initial pump pulse centered at $\omega_0$.

## 4. Discussion and conclusion

Firstly, we have shown that a pump short pulse built from the superposition of two beams having different topological charges nonlinearly propagates along a helical trajectory in the waveguide. Close to the critical power, the nonlinearly-induced formation of an ultra-broadband localized structure (pulse or shock front according to the dispersion regime) then takes place, thus generating the main axial spectral broadening. Besides, it also emits resonant radiations through intra- and intermodal properties (conical spectral broadening) satisfying the phase-matching condition described by Eq. 25, and whose linear superposition forms a spatiotemporal helicon wavepacket with its group velocity and rotating rate (i.e., distinct from characteristics of initial pump pulse due to the nonlinear effects which implies corrections $\delta k_1$ and $\delta k_l$). It appears that our analysis also unveils the intensity dependence of both group velocity and rotation rate of the ultra-broadband localized structure at the origin of the helicon wavepacket. One well-known physical effect at the origin of such a dependence of the group velocity is the self-steepening effect. It usually induces a clear optical shock appearing on the trailing edge of an ultrashort pulse when studied in the normal dispersion regime, which can also seed a conical wave of X-type [34,43]. In the present case, this effect is less noticeable but still present at the pulse splitting stage. The formation of an optical shock results from the velocity difference between the pulse peak and its tails due to the nonlinear dependence of the refractive index. In the same way, we observe such equivalent dynamics in the spatial domain, as shown in Figs. 2(e) and 3, featured by an azimuthal self-steepening of the intensity profiles on the trailing edge. The study of combined temporal and azimuthal shocks (i.e., a helical shock) then appears as a possible general mechanism responsible for the emergence of spatiotemporal helicon wavepackets. Mathematically, one can simply refer to the first-order approximation of the frequency-dependent nonlinearity $\frac{i\omega^2}{2\epsilon_0 c^2 K_z}$ in UPPE (see Eq. 22) by setting $K_z = K_0 + K_1(\omega - \omega_0) + K_l l$, in order to retrieve the origin of simultaneous self-steepening in both temporal and spatial domains. Indeed, this would imply in the space-time domain the simplified operator acting on the nonlinear term $\omega_0/n_0 c\ T$, where $T = \left[1 + i/\omega_0\ \partial_t + i\ K_l/K_0\ \partial_\theta\right]$ is the spatiotemporal self-steepening.

Secondly, it is worth mentioning that, while the above example here considered a superposition of $l = 0$ and $l = 1$ as initial conditions, the dynamics still remains the same if one uses another superposition of two others OAM modes. For instance, our simulations showed (data not shown) that using an initial superposition of $l = 0$ and $l = 2$ also leads to the emergence of a helicon wavepacket. In that case, however, the generated wavepacket is only composed of modes with even topological charges due to the selection rules imposed by the symmetry. More generally, close to the critical power, we have found that preparing the initial pump pulse as a superposition of multiple OAMs of low topological charges always leads to the emergence of spatiotemporal helicon wavepackets. This is due to the low dispersion of the propagation constant with respect to the topological charge $l$. Even prepared with multiple low-order OAMs, the pump pulse still rotates around the propagation axis without significant angular dispersion.

Finally, OAM fiber modes are known to provide an additional degree of freedom with which to control nonlinear wave propagation [44-45], but this also raises the usual challenge of mode coupling that OAM modes face when propagating in a conventional multimode fiber [36-37]. A relatively small difference between the effective refractive indices of two modes favors power exchange (i.e., coupling) during propagation. Recently, specially designed fibers (i.e., vortex fibers) have been developed to reduce the mutual mode coupling and support stable propagation of OAM modes over long distances [36-37]. We recall that our description and findings are valid in any waveguide with axisymmetric refractive index profile, so that we expect that future studies in specialty OAM waveguides will be successful. Note that the spontaneous generation of helicon wavepackets in the case of nonlinear propagation can be also observed on very short distances as shown previously.

In conclusion, we have presented a novel kind of propagation-invariant space-time wavepackets that have the particularity to rotate around both propagation and time axis. The invariance property of such electric fields is due to tight space-OAM-frequency correlations that compensate for dispersion and diffraction at the same time. The possibilities and interests of such helicon wavepackets are twofold. First, with the actual technologies of modal multiplexing and fiber fabrication, the experimental generation of such linear wavepackets in multimode fibers seems attainable. Beyond the fundamental interest of their observations, we strongly believe that potential applications in laser-matter manipulation could emerge from the generation of such wavepackets. Our approach could be also extended to more complex structured beams with time-varying OAM also known as self-torqued beams [46]. Second, we have shown that twirling wavepackets can naturally emerge from the nonlinear propagation of rather intense ultrashort pulses. As for conical waves emerging during filamentation in bulks, the formalism of helicon wavepackets brings a proper understanding of the phenomena taking place when intense pulses embedding different OAM beams propagate in bulks and waveguides. Given the current strong interest of both OAM-carrying beams and fiber modes, we expect that this work will stimulate further researches in the field. More specifically, we anticipate that space-time helicon wavepackets could emerge during the filamentation process in bulk media since Laguerre-Gauss beams are known to exhibit an OAM-dependent Gouy phase. Superimposed Laguerre-Gauss beams whose azimuthal mode index is correlated to their frequency are associated to helical structures for their phase and intensity profiles in space close to the waist [24,47]. As a consequence, such a field combined with our phase-matching could then generate space-time helicon wavepacket if its power exceeds the critical power necessary to trigger the filamentation process. If so, it suggests important applications in terms of processing of transparent dielectrics.

Finally, from a general point of view, our work provides a general approach to explore the dynamics of three-dimensional spiraling wavepackets with topological properties, the outlook of this topic going well beyond optics since being studied nowadays in various branches of physics such as acoustic spanners [48], polaritonics [49], plasma waves [50], and electron or particle beams [51].

**Funding.** Agence Nationale de la Recherche (EIPHI Graduate School, Contract No. ANR-17-EURE-0002, The Smile project; PIA2/ ISITE-BFC, contract ANR-15-IDEX-03, Breathing Light project). European Regional Development Fund.

**Acknowledgments.** The authors thank DNUM-CCUB (Université de Bourgogne) for HPC resources used for numerical simulations.

**Disclosures.** The authors declare no conflicts of interest.

**Data availability.** Data underlying the results presented in this paper are not publicly available at this time but may be obtained from the authors upon reasonable request.

**Supplemental document.** See Supplementary movie for supporting content.

## References

1.  A. Hansen, J. T. Schultz, and N. P. Bigelow, "Singular atom optics with spinor Bose-Einstein condensates," Optica **3**, 355-361 (2016).
2.  J. Wang, J-Y. Yang, I. M. Fazal, N. Ahmed, and Y. Yan, "Terabit free-space data transmission employing orbital angular momentum multiplexing," Nat. Photon. **6**, 488-496 (2012).
3.  Z. Xie, T. Lei, F. Li, H. Qiu, and Z. Zhang, "Ultra-broadband on-chip twisted light emitter for optical communications," Light. Sci. Appl. **7**, 18001 (2018).
4.  K. Dholakia and T. Čižmár, "Shaping the future of manipulation," Nat. Photon. **5**, 335-342 (2011).
5.  M. Woerdemann, C. Alpmann, M. Esseling, and C. Denz, "Advanced optical trapping by complex beam shaping," Laser Photon. Rev. **7**, 839-854 (2013).
6.  A. Sit, F. Bouchard, R. Fickler, J. Gagnon-Bischoff, and H. Larocque, "High-dimensional intracity quantum cryptography with structured photons," Optica **4**, 1006-1010 (2017).
7.  M. Erhard, R. Fickler, M. Krenn, and A. Zeilinger, "Twisted photons: New quantum perspectives in high dimensions," Light. Sci. Appl. **7**, 17146 (2018).
8.  A. Mathis, F. Courvoisier, L. Froehly, L. Furfaro, and M. Jacquot, "Micromachining along a curve: Femtosecond laser micromachining of curved profiles in diamond and silicon using accelerating beams," Appl. Phys. Lett. **101**, 071110 (2012).
9.  F. Courvoisier, R. Stoian, and A. Couairon, "Ultrafast laser micro-and nano-processing with nondiffracting and curved beams: invited paper for the section: hot topics in ultrafast lasers," Opt. Laser Technol. **80**, 125-137 (2016).
10. F. O. Fahrbach, P. Simon, and A. Rohrbach, "Microscopy with self-reconstructing beams," Nat. Photon. **4**, 780-785 (2010).
11. Q. Zhan, "Cylindrical vector beams: from mathematical concepts to applications," Adv. Opt. Photon. **1**, 1-57 (2009).
12. K. Liu, Y. Cheng, Y. Gao, X. Li, Y. Qin, and H. Wang, "Super-resolution radar imaging based on experimental OAM beams," Appl. Phys. Lett. **101**, 071110 (2017).
13. J. Nylk, K. McCluskey, M. A. Preciado, M. Mazilu, F. J. Gunn-Moore, S. Aggarwal, J. A. Tello, D. E. K. Ferrier, and K. Dholakia, "Light-sheet microscopy with attenuation-compensated propagation-invariant beams," Sci. Adv. **4**, eaar4817 (2018).
14. J. Durnin, J. J. Miceli, Jr., and J. H. Eberly, "Diffraction-free beams," Phys. Rev. Lett. **58**, 1499 (1987).
15. Z. Bouchal, "Nondiffracting optical beams: Physical properties, experiments, and applications," Czechoslov. J. Phys. **53**, 537-578 (2003).
16. J. C. Gutiérrez-Vega, M. D. Iturbe-Castillo, and S. Chávez-Cerda, "Alternative formulation for invariant optical fields: Mathieu beams," Opt. Lett. **25**, 1493-1495 (2000).
17. M. A. Bandres and B. M. Rodríguez-Lara, "Nondiffracting accelerating waves: Weber waves and parabolic momentum," New J. Phys. **15**, 013054 (2003).
18. G. A. Siviloglou, J. Broky, A. Dogariu, and D. N. Christodoulides, "Observation of accelerating Airy beams," Phys. Rev. Lett. **99**, 213901 (2007).

19. L. Allen, M. J. Padgett, and M. Babiker, "IV. The Orbital Angular Momentum of Light," in *Progress in Optics*, vol. 39 E. Wolf, ed. (Elsevier, 1999), pp. 291–372.
20. Y. Shen, X. Wang, Z. Xie, C. Min, X. Fu, Q. Liu, M. Gong, and X. Yuan, "Optical vortices 30 years on: OAM manipulation from topological charge to multiple singularities," Light. Sci. Appl. **8**, 90 (2019).
21. S. Chávez-Cerda, G. S. McDonald, and G. H. C. New, "Nondiffracting beams: travelling, standing, rotating and spiral waves," Opt. Commun. **123**, 225-233 (1996).
22. S-H. Lee, Y. Roichman, and D. G. Grier, "Optical solenoid beams," Opt. Express **18**, 6988-6993 (2010).
23. C. Vetter, T. Eichelkraut, M. Ornigotti, and A. Szameit, "Generalized radially self-accelerating Helicon beams," Phys. Rev. Lett. **113**, 183901 (2014).
24. Z. Zhao, H. Song, R. Zhang, K. Pang, C. Liu, H. Song, A. Almaiman, K. Manukyan, H. Zhou, B. Lynn, R. W. Boyd, M. Tur, and A. E. Willner, "Dynamic spatiotemporal beams that combine two independent and controllable orbital-angular-momenta using multiple optical-frequency-comb lines," Nat. Commun. **11**, 4099 (2020).
25. C. Schulze, F. S. Roux, A. Dudley, R. Rop, M. Duparré, and A. Forbes, "Accelerated rotation with orbital angular momentum modes," Phys. Rev. A **91**, 043821 (2015).
26. A. Brimis, K. G. Makris, and D. G. Papazoglou, "Tornado waves," Opt. Lett. **45**, 280-283 (2020).
27. D. B. Ruffner and D. G. Grier, "Optical Conveyors: A Class of Active Tractor Beams," Phys. Rev. Lett. **109**, 163903 (2012).
28. W. Ding, T. Zhu, L-M. Zhou, and C-W. Qiu, "Photonic tractor beams: a review," Adv. Photon. **1**, 024001 (2019).
29. J. N. Brittingham, "Focus waves modes in homogeneous Maxwell's equations: Transverse electric mode," J. Appl. Phys. **54**, 1179 (1998).
30. L. Mackinnon, "A nondispersive de Broglie wave packet," Found. Phys. **8**, 157-176 (1978).
31. H. Sõnajalg, M. Rätsep, and P. Saari, "Demonstration of the Bessel-X pulse propagating with strong lateral and longitudinal localization in a dispersive medium," Opt. Lett. **22**, 310-312 (1997).
32. H. E. Kondakci and A. F. Abouraddy, "Diffraction-free space–time light sheets," Nat. Photon. **11**, 733-740 (2017).
33. C. Conti, S. Trillo, P. Di Trapani, G. Valiulis, A. Piskarskas, O. Jedrkiewicz, and J. Trull, "Nonlinear Electromagnetic X Waves," Phys. Rev. Lett. **90**, 170406 (2003).
34. B. Kibler and P. Béjot, "Discretized conical waves in multimode optical fibers," Phys. Rev. Lett. **126**, 023902 (2021).
35. J. Andreasen and M. Kolesik, "Nonlinear propagation of light in structured media: Generalized unidirectional pulse propagation equations," Phys. Rev. E **86**, 036706 (2012).
36. S. Ramachandran and P. Kristensen, "Optical vortices in fiber," Nanophotonics **2**, 455-474 (2013).
37. Z. Ma and S. Ramachandran, "Propagation stability in optical fibers: role of path memory and angular momentum," Nanophotonics **10**, 209-224 (2021).
38. M. A. Eftekhar, L. G. Wright, M. S. Mills, M. Kolesik, R. Amezcua Correa, F. W. Wise, and D. N. Christodoulides, "Versatile supercontinuum generation in parabolic multimode optical fibers," Opt. Express **25**, 9078-9087 (2017).
39. O. Tzang, A. M. Caravaca-Aguirre, K. Wagner, and R. Piestun, "Adaptive wavefront shaping for controlling nonlinear multimode interactions in optical fibres," Nat. Photon. **12**, 368-374 (2018).
40. E. Deliancourt, M. Fabert, A. Tonello, K. Krupa, A. Desfarges-Berthelemot, V. Kermene, G. Millot, A. Barthélémy, S. Wabnitz, and V. Couderc, "Wavefront shaping for optimized many-mode Kerr beam self-cleaning in graded-index multimode fiber," Opt. Express **27**, 17311-17321 (2019).
41. P. Béjot, "Multimodal unidirectional pulse propagation equation," Phys. Rev. E **99**, 032217 (2019).
42. K. Tarnowski, S. Majchrowska, P. Béjot, and B. Kibler, "Numerical modelings of ultrashort pulse propagation and conical emission in multimode optical fibers," J. Opt. Soc. Am. B **38**, 732-742 (2021).
43. M. Kolesik, E. M. Wright, and J. V. Moloney, "Interpretation of the spectrally resolved far field of femtosecond pulses propagating in bulk nonlinear dispersive media," Opt. Express **13**, 10729-10741 (2005).
44. K. Rottwitt, J. G. Koefoed, K. Ingerslev, and P. Kristensen, "Intermodal Raman amplification of OAM fiber modes," APL Photonics **4**, 030802 (2019).
45. X. Liu, E. N. Christensen, K. Rottwitt, and S. Ramachandran, "Nonlinear four-wave mixing with enhanced diversity and selectivity via spin and orbital angular momentum conservation," APL Photonics **5**, 010802 (2020).
46. L. Rego, K. M. Dorney, N. J. Brooks, Q. L. Nguyen, C-T. Liao, J. San Román, D. E. Couch, A. Liu, E. Pisanty, M. Lewenstein, L. Plaja, H. C. Kapteyn, M. M. Murnane, and C. Hernández-García, "Generation of extreme-ultraviolet beams with time-varying orbital angular momentum," Science **364**, eaaw9486 (2019).
47. G. Pariente and F. Quéré, "Spatio-temporal light springs: extended encoding of orbital angular momentum in ultrashort pulses," Opt. Lett. **40**, 2037-2040 (2015).
48. T. Brunet, J-L. Thomas, R. Marchiano, and F. Coulouvrat, "Experimental observation of azimuthal shock waves on nonlinear acoustical vortices," New J. Phys. **11**, 013002 (2009).
49. L. Dominici, D. Colas, A. Gianfrate, A. Rahmani, V. Ardizzone, D. Ballarini, M. De Giorgi, G. Gigli, F. P. Laussy, D. Sanvitto, and N. Voronova, "Full-Bloch beams and ultrafast Rabi-rotating vortices," Phys. Rev. Res. **3**, 013007 (2021).

50. J. Vieira, J. T. Mendonça, and F. Quéré, "Optical control of the topology of laser-plasma accelerators," Phys. Rev. Lett. **121**, 054801 (2018).
51. J. Pierce, J. Webste, H. Larocque, E. Karimi, B. McMorran, and A. Forbes, "Coiling free electron matter waves," New J. Phys. **21**, 043018 (2019).